\documentclass[epj]{svjour}
\usepackage{amsmath,graphicx,breqn,bm,rotating}

\def\py{p_{v, \mathrm{y}}}

\def\pb{p_{\mathrm{b}}}
\def\mt{m_{\mathrm{T}}}
\def\mtp{m'_{\mathrm{T}}}

\def\met{E^{\mathrm{miss}}_{\mathrm{T}}}

\begin{document}
\title{Transformation properties of the transverse mass under transverse Lorentz boosts at hadron colliders}
\titlerunning{Transformation properties of the transverse mass}
\author{Daniel R. Tovey
}                     
%
%
\institute{Department of Physics and Astronomy, University of Sheffield, Hounsfield Road, Sheffield S3 7RH, UK}
\date{Received: date / Revised version: date}
%
\abstract{
The transverse mass of semi-invisibly decaying particles, calculated from the transverse momenta of their decay products, has been used in a broad range of searches and measurements at hadron colliders, such as the LHC. This variable is invariant by construction under Lorentz boosts purely in the longitudinal (beam) direction, thereby minimising sensitivity to fluctuations in the fractions of the proton momentum carried by the colliding partons. In this paper we examine, by contrast, the properties of the transverse mass under boosts with a component also in the transverse plane perpendicular to the beam direction. We show that this variable is invariant under such boosts in cases where the boost is purely transverse and (a) the momenta of the decay products are confined to the transverse plane in the rest frame of the parent particle and/or (b) the transverse momenta of the decay products are perpendicular to the boost direction. We discuss the transformation properties of the transverse mass in the case of combined transverse and longitudinal boosts and identify the criteria under which the transverse mass in the laboratory frame can equal the rest mass of the parent particle, irrespective of its value in the rest frame of the parent.%
} 
\maketitle
\section{Introduction}
\label{sec1}
A key challenge at hadron colliders such as the Large Hadron Collider (LHC) \cite{Evans:2008zzb} is the identification and measurement of particles producing both visible and invisible decay products. Such semi-invisibly decaying particles include the Standard Model $W$ boson, Higgs boson (e.g. when decaying to $\ell^+\ell^-\nu\bar{\nu}$ via $ZZ^*$ or $W^+W^-$) and $\tau$ lepton, and many possible new beyond Standard Model states such as $W'$ bosons \cite{Aaboud:2017efa,Sirunyan:2018mpc,Sirunyan:2018lbg,Tanabashi:2018oca:wprime}, dark matter mediator particles \cite{Aaboud:2018zpr,Boveia:2018yeb} and Vector-Like Quarks \cite{Aaboud:2017zfn,Tanabashi:2018oca:vlq}. Kinematic variables constructed from the decay products of such particles which take values strongly correlated with their mass provide effective tools for both identifying their presence in events and measuring their masses. Variables defined using momentum components in the transverse plane make allowance for the lack of knowledge of the momenta of invisible particles in the beam direction.

A simple example of such a variable is just the net transverse momentum of the visible decay product(s) of the parent particle. The distribution of this variable displays a Jacobian peak in the parent rest frame at half the parent mass for massless decay products. The transverse momenta of visible decay products are invariant under Lorentz boosts of the parent rest frame along the beam direction. Hence the unknown value of this boost in each event, arising from the varying fractions of the proton momenta carried by the colliding partons, does not affect the position or shape of the peak. However, transverse momenta are {\it not} invariant under boosts {\it in} the transverse plane, generated, for instance, by recoil of the parent against hadrons in the final state of the hard process. For this reason the position and shape of the peak can be strongly affected by the physics of the hard process generating the hadronic recoil, and identification efficiencies and mass measurements using visible transverse momenta can be sensitive to recoil modelling uncertainties. Nevertheless, the use of transverse momenta of visible decay products provides an important route to measuring the masses of semi-invisibly decaying particles, and offers the benefit of insensitivity to potentially significant systematic uncertainties arising from experimental measurement of the net momenta of invisible decay products via the missing transverse energy ($\met$). The most recent measurement of the mass of the $W$ boson, by the ATLAS collaboration at the LHC, relies primarily upon this technique \cite{Aaboud:2017svj}. There has also been recent interest in the use of the energies of visible decay products with similar motivation \cite{Agashe:2012bn,Bianchini:2019iey}.

\begin{figure}[htb]
\begin{center}
\includegraphics[width=0.51\textwidth]{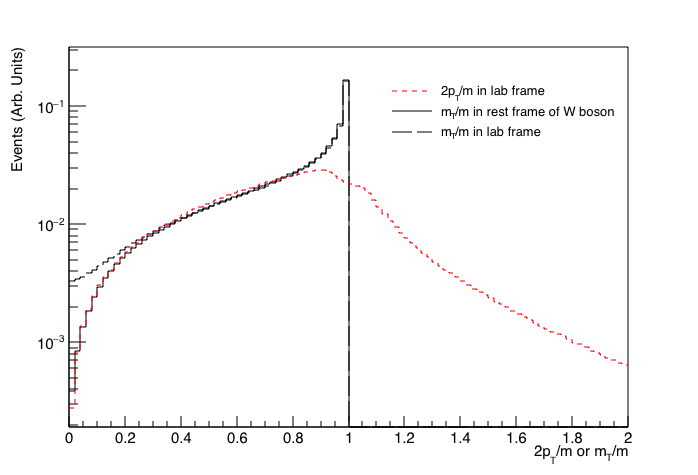}
\includegraphics[width=0.477\textwidth]{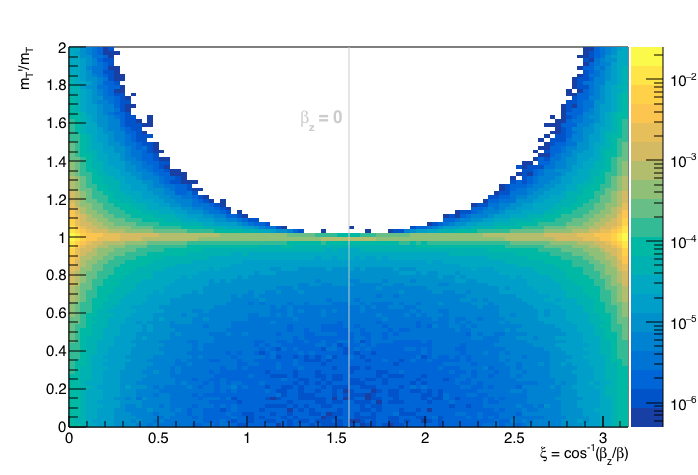}
\end{center}
\caption{\label{fig:mt} Top: Distributions of transverse mass values in the parent rest frame and in the laboratory frame for a $W$ boson decaying to a charged lepton and neutrino and recoiling against a hadronic system with transverse momentum of at least 5 GeV. Also shown (red/light dashed) is the equivalent distribution of charged lepton transverse momenta measured in the laboratory frame. Events were simulated with {\tt PYTHIA v.8.240} \cite{Sjostrand:2014zea}, specifying the $qg\rightarrow Wq$ process with ISR, FSR and hadronisation disabled. The transverse mass has been normalised to the per-event mass of the $W$ boson $m$ for each event to illustrate the form of the Jacobian peak. Bottom: The distribution of the ratio of the transverse mass values measured in the laboratory and $W$ boson frames plotted as a function of the angle $\xi$ between the boost and the beam-line ($\hat{z}$ direction) defined by $\cos\xi \equiv(\beta_{\mathrm{z}}/\beta)$.}
\end{figure}

An alternative to the direct use of transverse momenta is provided by the transverse mass ($\mt$) \cite{vanNeerven:1982mz,Arnison:1983rp,Smith:1983aa}, defined as the invariant mass of the visible and invisible decay products calculated assuming that each has zero net momentum in the beam direction. Distributions of $\mt$ also display a Jacobian peak, at the mass of the parent particle. Values of $\mt$ are also invariant under Lorentz boosts in the beam direction, as they depend purely upon the transverse momenta of the decay products (including $\met$ for invisible particles). The sensitivity of the position and shape of the peak to transverse boosts of the parent is furthermore reduced in comparison with the visible transverse momentum distribution discussed above (see Figure~\ref{fig:mt}~(top)), due to the similarity of the definition of $\mt$ to that of the invariant mass. However, $\mt$ is not completely invariant under purely transverse boosts \cite{Smith:1983aa} or indeed under more general combined transverse and longitudinal boosts (see Figure~\ref{fig:mt}~(bottom)). In this paper we will examine in detail the transformation properties of this variable in these cases. For simplicity the paper will focus on the special case of a singly-produced massive parent ($\delta$) decaying to two massless visible ($v$) and invisible ($\chi$) daughter particles. 

The paper begins by defining the coordinate system and kinematic quantities to be used (Section~\ref{sec2}). There follows (Section~\ref{sec3}) an analysis of the values of $\mt$ measured in the laboratory frame in the presence of a purely transverse Lorentz boost of the semi-invisibly decaying parent particle arising from its transverse recoil.
Next, the values of $\mt$ measured in the laboratory frame in the presence of a more general combined transverse and longitudinal Lorentz boost are considered (Section~\ref{sec4}). The paper finishes with a discussion of these findings (Section~\ref{sec5}) and conclusions (Section~\ref{sec6}).

\section{Coordinate system and kinematic variables}
\label{sec2}
To simplify the analysis we assume initially that the momenta of both $\delta$ and the system against which it recoils lie in the transverse plane of the laboratory frame -- in other words we assume that the complete final state system of the hard process is at rest in the beam direction (defined as $\hat{\mathrm{z}}$). We later relax this assumption and investigate the non-trivial consequences that this has for $\mt$. We assume that the masses of the visible and invisible particles can be neglected both in the definition of the transverse mass and in the analysis that follows. 

In the rest frame of $\delta$ the visible and invisible decay products have momenta of magnitude $p_0 = m/2$. Here, $m$ is the per-event mass of $\delta$, sampled from a Breit-Wigner distribution of mean $m_{\delta}$ and width $\Gamma_{\delta}$. The particle $\delta$ recoils in the laboratory frame against a system of net momentum $\mathbf{p}_{\mathrm{b}}$. We define the decay angle relative to the beam direction ($\hat{\mathrm{z}}$), in the rest frame of $\delta$, as $\theta$ such that $\theta=\pi/2$ corresponds to decays confined to the transverse plane while $\theta=0$ or $\pi$ correspond to decays parallel to the beam direction. We define the decay angle $\phi$ in the transverse plane of the rest frame of $\delta$, relative to the transverse momentum $\mathbf{p}_{\delta,\mathrm{T}}$ of $\delta$, and hence to the Lorentz boost, such that $\phi=0$ corresponds to a visible (invisible) daughter momentum parallel (anti-parallel) to $\mathbf{p}_{\delta,\mathrm{T}}$. We define the positive $\hat{\mathrm{x}}$ direction as being parallel to the transverse momentum of $\delta$ and hence the boost direction in the transverse plane, and the positive $\hat{\mathrm{y}}$ direction in the transverse plane such that the visible daughter has positive $\hat{\mathrm{y}}$ momentum component $\py$.

The recoil of $\delta$ leads to a boost of its decay products with velocity ${\bm \beta} = \mathbf{p}_{\mathrm{b}}/{(\pb^2+m^2)}$ in the laboratory frame. The relativistic boost factor is $\gamma = 1/\sqrt{1-\beta^2}=\sqrt{\pb^2+m^2}/m$. We denote the rest frame of $\delta$ as the unprimed frame, and the detector / laboratory frame as the primed frame.

The invariant mass of the daughters, $v$ and $\chi$, is equal to the per-event mass $m$ of $\delta$ and can be defined as follows:
\begin{eqnarray}
m_{v\chi}^2 &=& (E_v+E_\chi)^2 - (\mathbf{p}_v+\mathbf{p}_\chi)^2\cr
&=& m_v^2+m_\chi^2+2\left(E_vE_\chi-p_{v, \mathrm{x}}p_{\chi, \mathrm{x}}-p_{v, \mathrm{y}}p_{\chi, \mathrm{y}}\right.\cr&&\left.\hspace{40mm}-p_{v, \mathrm{z}}p_{\chi, \mathrm{z}}\right)\cr
&=& m_v^2+m_\chi^2+2\left(E_{v, \mathrm{T}}E_{\chi, \mathrm{T}}\cosh\Delta\eta-p_{v, \mathrm{x}}p_{\chi, \mathrm{x}}\right.\cr&&\left.\hspace{40mm}-p_{v, \mathrm{y}}p_{\chi, \mathrm{y}}\right),
\label{eq:minv}
\end{eqnarray}
where $E_{i, \mathrm{T}}^2\equiv p_{i, \mathrm{T}}^2+m_i^2$ and $p_{i, \mathrm{T}}^2\equiv p_{i, \mathrm{x}}^2+p_{i, \mathrm{y}}^2$ for particle $i$. $\Delta\eta$ is the difference in rapidity for particles $v$ and $\chi$. The rapidity is defined by
\begin{equation}
\eta \equiv \frac{1}{2}\ln\left(\frac{E_i+p_{i,\mathrm{z}}}{E_i-p_{i,\mathrm{z}}}\right)
\end{equation}
for particle $i$, and reduces to $\eta=-\ln(\tan(\theta/2))$ for massless $v$ and $\chi$. Differences in rapidity $\Delta\eta$ are invariant under Lorentz boosts in the longitudinal ($\hat{\mathrm{z}}$) direction, but are not invariant under boosts in other directions.

As noted in Section~\ref{sec1}, the transverse mass $\mt$ of the $v\chi$ system is defined as the invariant mass under the assumption that components of the momenta of $v$ and $\chi$ in the beam direction are zero. This is equivalent to setting $\Delta\eta=0$ in Eqn.~(\ref{eq:minv}):
\begin{eqnarray}
\mt^2 &=& \left(E_{v, \mathrm{T}}+E_{\chi, \mathrm{T}}\right)^2 - \left(\mathbf{p}_{v, \mathrm{T}}+\mathbf{p}_{\chi, \mathrm{T}}\right)^2\cr
&=& m_v^2+m_\chi^2+2\left(E_{v, \mathrm{T}}E_{\chi, \mathrm{T}}-p_{v, \mathrm{x}}p_{\chi, \mathrm{x}}-p_{v, \mathrm{y}}p_{\chi, \mathrm{y}}\right)\cr
&=& 2\left(p_{v, \mathrm{T}}p_{\chi, \mathrm{T}}-p_{v, \mathrm{x}}p_{\chi, \mathrm{x}}-p_{v, \mathrm{y}}p_{\chi, \mathrm{y}}\right),
\label{eqn:mt}
\end{eqnarray}
where we have required $m_v = m_\chi = 0$ in the last line. In general $\mt\le m$, because $\cosh\Delta\eta\ge 1$ and hence $\mt\le m_{v\chi}$, as can be seen from the identity:
\begin{equation}
\mt^2 = m_{v\chi}^2 +2E_{v, \mathrm{T}}E_{\chi, \mathrm{T}}(1-\cosh\Delta\eta).
\label{eqn:mmt}
\end{equation}

The transverse mass is invariant under arbitrary longitudinal boosts of the laboratory transverse plane along the beam direction, because it is defined exclusively in terms of momentum components measured in the transverse plane.

\section{Transformation properties of $\mt$ under purely transverse Lorentz boosts}
\label{sec3}

Before considering the more general case of a combined transverse and longitudinal Lorentz boost of $\delta$ in the laboratory frame we consider first the simpler case of a purely transverse boost, for which ${\bm\beta}$ lies in the transverse plane and hence ${\bm\beta}=(\beta_{\mathrm{x}},0,0)$ and $\beta = \beta_{\mathrm{x}}$. This case was also considered in detail in Ref.~\cite{Smith:1983aa}.

\subsection{Transverse mass in the laboratory frame}
With the definitions from Section~\ref{sec2} we can obtain the momentum components of $v$ and $\chi$ in the laboratory (primed) frame from a Lorentz transformation of the unprimed quantities:\\
\noindent\begin{minipage}{.5\linewidth}
\begin{eqnarray}
p_{v, \mathrm{x}}' &=& \gamma (p_{v, \mathrm{x}} + \beta E_{v})\cr
&=& p_0\gamma(\sin\theta\cos\phi+\beta), \cr
p_{\chi, \mathrm{x}}' &=& \gamma (p_{\chi, \mathrm{x}} + \beta E_{\chi})\cr
&=& p_0\gamma(-\sin\theta\cos\phi+\beta), \nonumber
\end{eqnarray}
\end{minipage}%
\begin{minipage}{.5\linewidth}
\begin{eqnarray}
p_{v, \mathrm{y}}' &=& p_{v, \mathrm{y}} \cr
&=& p_0\sin\theta\sin\phi, \cr
p_{\chi, \mathrm{y}}' &=& p_{\chi, \mathrm{y}} \cr
&=& -p_0\sin\theta\sin\phi.
\label{eqn:boostx}
\end{eqnarray}
\end{minipage}\newline

We want to calculate the value of the transverse mass in the laboratory frame ($\mtp$) and so we must use the primed momentum components. Substituting with the above Lorentz transformations and performing some algebra we obtain the following:
\begin{eqnarray}
{\mtp}^2 &=& 2\left(p'_{v, \mathrm{T}}p'_{\chi, \mathrm{T}}-p'_{v, \mathrm{x}}p'_{\chi, \mathrm{x}}-p'_{v, \mathrm{y}}p'_{\chi, \mathrm{y}}\right),\cr
&=&\frac{m^2}{2}\Biggl(\sin^2\theta+(1-\gamma^2)(1-\sin^2\theta\cos^2\phi)\Biggr.\cr&&\Biggl.+\Bigl|\left((\gamma^2(1-\sin^2\theta\cos^2\phi)-(1-\sin^2\theta\sin^2\phi))^2\right.\Bigr.\Biggr.\cr&&\Biggl.\Bigl.\left.+4\gamma^2\sin^4\theta\cos^2\phi\sin^2\phi\right)^{1/2}\Bigr|\Biggr).\cr
\label{eqn:master}
\end{eqnarray}

\subsection{Special cases}
\label{sec32}
We can now examine the values of $\mtp$ for several special cases.
\begin{itemize}

\item {\bf $\gamma=1$, any $\phi$, any $\theta$:} In this case $\delta$ is at rest in the laboratory transverse plane. $\mtp$ is then just given by $\mtp = m\sin\theta$, independent of $\phi$. 

\item {\bf $\theta=\pi/2$, any $\phi$, any $\gamma$:} In this case $\delta$ decays in the transverse plane. The rapidity difference for $v$ and $\chi$ vanishes and hence $\mtp = m$, independent of $\phi$ and $\gamma$. $\mtp$ thus measures $m$ independently of the magnitude or direction of the recoil. This special case is the origin of an invariant population of events in the Jacobian peak of the $\mtp$ distribution for purely transverse boosts. 

\item {\bf $\phi=\pi/2$, any $\theta$, any $\gamma$:} In this case $\delta$ decays in its rest frame perpendicularly to its direction of motion. The factors of $\gamma$ cancel from Eqn.~(\ref{eqn:master}) and we obtain $\mtp = m\sin\theta\le m$, independently of $\gamma$. Such events with $\theta\neq \pi/2$ therefore give rise to an invariant tail of events in the $\mtp$ distribution below the Jacobian peak at $\mtp=m$, for purely transverse boosts.

\item {\bf $\gamma> 1$, $\phi=0$ or $\pi$, and $\theta\neq \pi/2$:} In this case $\mtp$ is given by
\begin{equation}
\mtp = m\sqrt{1-\gamma^2\cos^2\theta},
\end{equation}
for $\beta<\sin\theta$, or equivalently $\gamma<1/\cos\theta$. When $\gamma\geq 1/\cos\theta$ the direction of the momentum of $\chi$ (or $v$ for $\phi=\pi$) in the transverse plane changes sign and becomes parallel to that of $v$ (or $\chi$). This causes $\mtp$ to become identically zero when the decay products are massless.

\item {\bf $\gamma\rightarrow\infty$:} In this case the limit of Eqn.~(\ref{eqn:master}) is given by:
\begin{eqnarray}
\mtp &=& \frac{m\sin\theta\sin\phi}{\sqrt{1-\sin^2\theta\cos^2\phi}}\cr
&=& \frac{m\ p_{v, \mathrm{y}}}{\sqrt{p_{v, \mathrm{y}}^2 + p_{v, \mathrm{z}}^2}}\cr
&=& m\sin\psi, 
\label{eqn:lim}
\end{eqnarray}
where $\psi$ is the visible daughter decay angle relative to the beam direction projected into the $\hat{\mathrm{y}}-\hat{\mathrm{z}}$ plane. Highly boosted events therefore acquire values of $\mtp$ which are always less than that measured in the parent's rest frame.

\end{itemize}

\begin{figure}[htb]
\begin{center}
\includegraphics[width=0.5\textwidth]{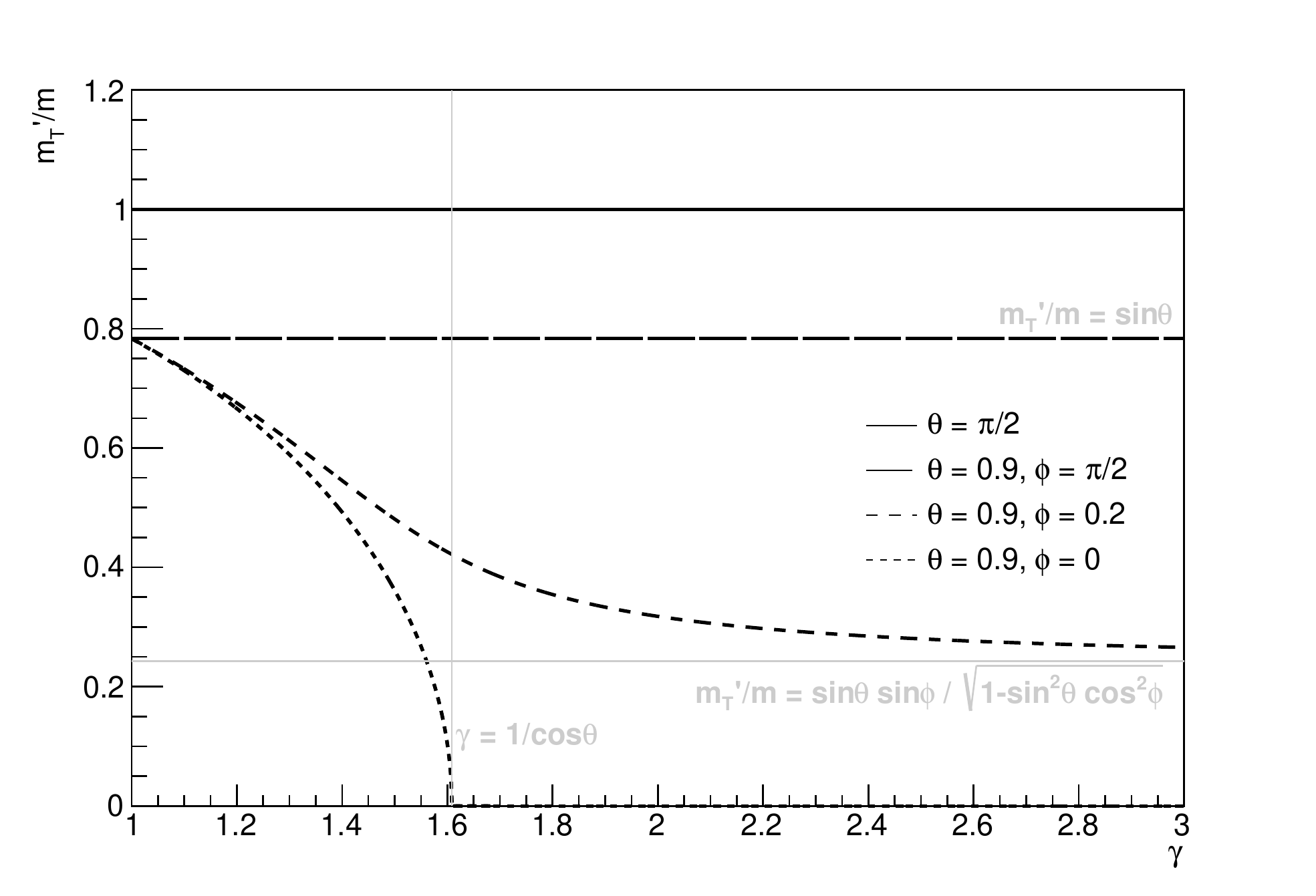}
\end{center}
\caption{\label{fig2}Values of $\mtp$ normalised to $m$ plotted as a function of $\gamma$ for different values of $\theta$ and $\phi$. The value of $\mtp/m$ at $\gamma=1$ for $\theta=0.9$ is $\sin\theta=0.783$. The maximum value of $\gamma$ for $\theta=0.9$ and $\phi=0$ for which $\mtp$ is non-zero is $\gamma=1/\cos\theta=1.609$. The curve in this case has equation $\mtp/m=\sqrt{1-0.386\gamma^2}$. The limiting value of $\mtp/m$ for $\theta=0.9$ and $\phi=0.2$, given by Eqn.~(\ref{eqn:lim}), is 0.243.}
\end{figure}

These cases are illustrated in Figure~\ref{fig2} for various values of $\theta$ and $\phi$. Excepting the above special cases, the general dependence of $\mtp$ on $\gamma$ resembles that observed for $\theta=0.9$ and $\phi=0.2$ in the figure. In general the shape of the $\mtp$ distribution is affected by the recoil of $\delta$, and $\mtp\le\mt\leq m$ (as noted also in Refs.~\cite{Smith:1983aa,Nachman:2016qyc}). The constraint that $\mtp/\mt\leq 1$ can be observed in Figure~\ref{fig:mt} (bottom) when $\xi=\pi/2$ and hence $\beta_{\mathrm{z}}=0$.

\section{Transformation properties of $\mt$ under combined transverse and longitudinal Lorentz boosts}
\label{sec4}

We now consider the more general case of a combined transverse and longitudinal Lorentz boost of $\delta$ in the laboratory frame. In this case, and with the definition that the boost points in the $+\hat{\mathrm{x}}$ direction in the transverse plane, we have ${\bm\beta} = (\beta_{\mathrm{x}}, 0, \beta_{\mathrm{z}})$ and $\beta=\sqrt{\beta_{\mathrm{x}}^2+\beta_{\mathrm{z}}^2}$. We define the angle of the boost relative to the beam direction as $\xi$ such that $\tan\xi \equiv(\beta_{\mathrm{x}}/\beta_{\mathrm{z}})$ or equivalently $\cos\xi \equiv(\beta_{\mathrm{z}}/\beta)$.

\subsection{Transverse mass in the laboratory frame}
\label{sec41}
With the definitions from Section~\ref{sec2} we can again obtain the momentum components of $v$ and $\chi$ in the laboratory (primed) frame from a Lorentz transformation of the unprimed quantities, using now the expression for a boost in an arbitrary non-axial direction:
\begin{eqnarray}
p_{v, \mathrm{x}}' &=& \gamma\beta_{\mathrm{x}} E_v + p_{v, \mathrm{x}} + (\gamma-1)\frac{\beta_{\mathrm{x}}}{\beta^2} (\beta_{\mathrm{x}} p_{v, \mathrm{x}} + \beta_{\mathrm{z}}  p_{v, \mathrm{z}})\cr
&=& p_0\Bigl(\gamma\beta_{\mathrm{x}}+\sin\theta\cos\phi  \Bigr.\cr&&\Bigl.\hspace{10mm} +(\gamma-1)\frac{\beta_{\mathrm{x}}}{\beta^2} (\beta_{\mathrm{x}} \sin\theta\cos\phi + \beta_{\mathrm{z}}  \cos\theta)\Bigr)\cr
p_{v, \mathrm{y}}' &=& p_{v, \mathrm{y}} \cr
&=& p_0\sin\theta\sin\phi, \cr
p_{\chi, \mathrm{x}}' &=& \gamma\beta_{\mathrm{x}} E_\chi + p_{\chi, \mathrm{x}} + (\gamma-1)\frac{\beta_{\mathrm{x}}}{\beta^2} (\beta_{\mathrm{x}} p_{\chi, \mathrm{x}} + \beta_{\mathrm{z}}  p_{\chi, \mathrm{z}})\cr
&=& p_0\Bigl(\gamma\beta_{\mathrm{x}}-\sin\theta\cos\phi \Bigr.\cr&&\Bigl.\hspace{10mm} -(\gamma-1)\frac{\beta_{\mathrm{x}}}{\beta^2} (\beta_{\mathrm{x}} \sin\theta\cos\phi  + \beta_{\mathrm{z}}  \cos\theta)\Bigr)\cr
p_{\chi, \mathrm{y}}' &=& p_{\chi, \mathrm{y}} \cr
&=& -p_0\sin\theta\sin\phi.
\label{eqn:boostxz}
\end{eqnarray}

\begin{figure}[htb]
\begin{center}
\includegraphics[width=0.5\textwidth]{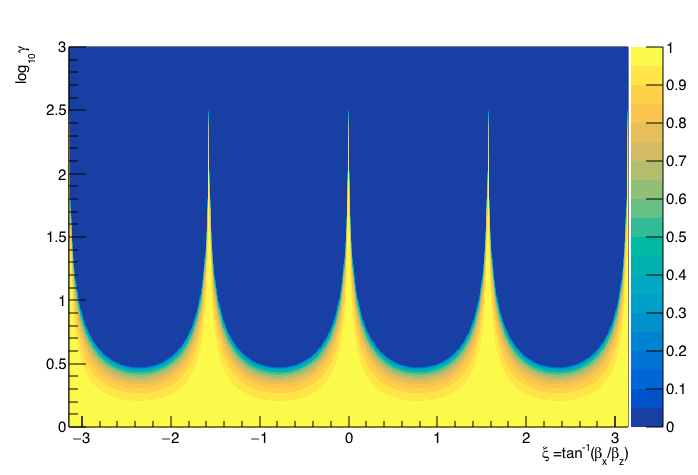}
\includegraphics[width=0.5\textwidth]{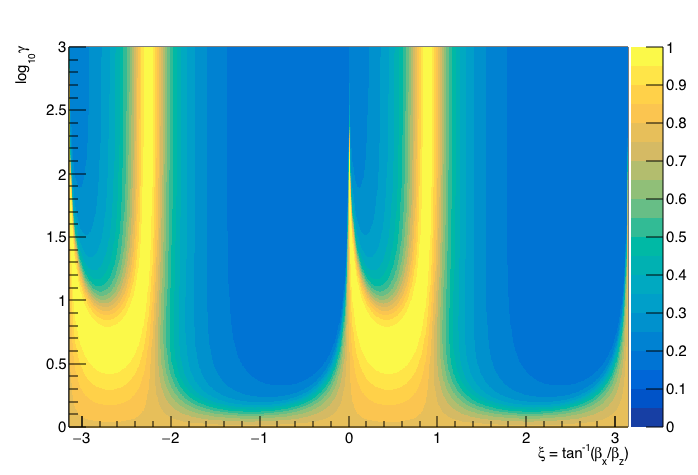}
\end{center}
\caption{\label{fig:gamxi}Values of $\mtp$ normalised to $m$ plotted as a function of $\log_{10}\gamma$ and $\xi$ for $\theta=\pi/2$ and $\phi=0$ (top) and $\theta=0.9$ and $\phi=0.2$ (bottom). In the top figure the boundary of the dark blue region, in which $\mtp=0$, is given by Eqn.~(\ref{eqn:gamlim}). With the values $\theta$ and $\phi$ used in the bottom figure, the value of $\mt/m$ in the $\delta$ rest frame is equal to $\sin\theta=0.783$.}
\end{figure}
Substituting these primed momenta into Eqn.~(\ref{eqn:mt}) we can obtain an expression for $\mtp$ similar to that in Eqn.~(\ref{eqn:master}) for purely transverse boosts. We do not provide its specific analytical form here, however we have plotted its dependence upon $\gamma$ and $\xi$ for two representative cases for $\theta$ and $\phi$ in Figure~\ref{fig:gamxi}. It is immediately noticeable in this figure that, in contrast to the case of purely transverse boosts, $\mtp$ can exceed $\mt$ (which is given by $m\sin\theta=0.783$ $m$ in this case), as observed also in Figure~\ref{fig:mt} (bottom) for $\xi\neq\pi/2$, and indeed can saturate the bound $\mtp\le m$.

\subsection{Special cases}
\label{sec42}

We can now again examine the values of $\mtp$ for several special cases.

\begin{figure}[htb]
\begin{center}
\includegraphics[width=0.47\textwidth]{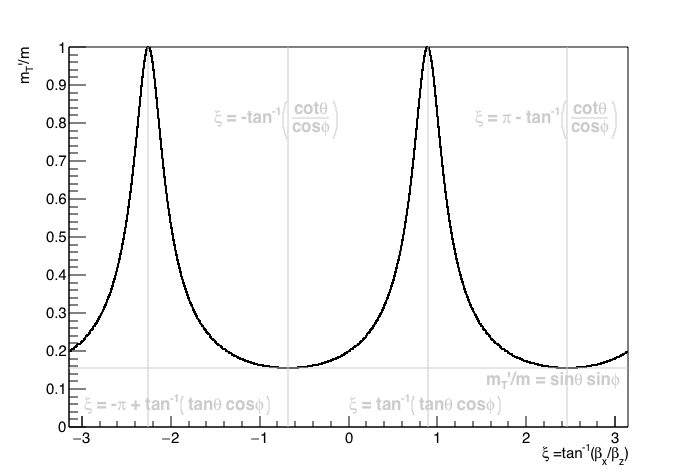}
\includegraphics[width=0.47\textwidth]{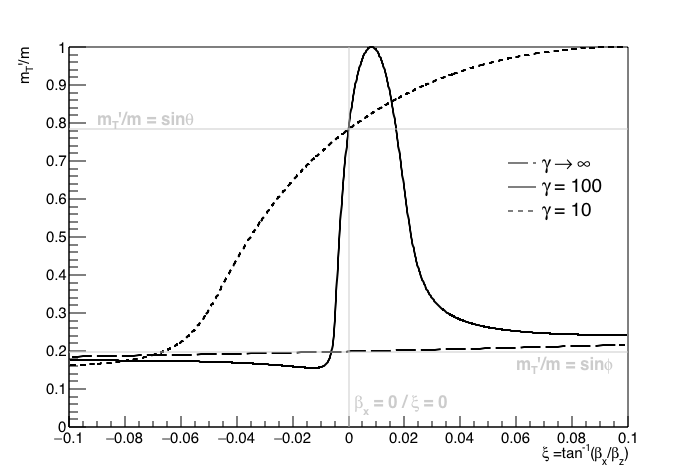}
\end{center}
\caption{\label{fig:gaminfty}Top: Values of $\mtp$ normalised to $m$ plotted as a function of $\xi$ for $\theta=0.9$ and $\phi=0.2$ for $\gamma\rightarrow\infty$. The positions of the maxima and minima given by Eqns.~\ref{eq:mtmax} and~\ref{eq:mtmin} are denoted by light grey lines, as is the minimum value of $\mtp/m = \sin\theta\sin\phi = 0.156$. Bottom: An enlargement of the top panel around $\xi=0$ (i.e. $\beta_{\mathrm{x}}=0$) with curves for several other large, but finite, values of $\gamma$ also shown. The two limiting values of $\mtp/m$ are shown -- namely $\mtp/m=\sin\phi=0.199$ when $\gamma\rightarrow\infty$ and $\mtp/m=\sin\theta=0.783$ when $\beta_{\mathrm{x}}=0$.}
\end{figure}

\begin{itemize}

\item {\bf $\theta=\pi/2$ and $\phi=\pi/2$, any $\gamma$:} In this case $\Delta\eta=0$ in the $\delta$ rest frame and is invariant under the boost, because the momenta of $v$ and $\chi$ are perpendicular to the boost direction. Consequently $\mtp=\mt=m$.

\item {\bf $\theta=\pi/2$ and $\phi\neq\pi/2$, any $\gamma$:} In this case $\Delta\eta=0$ in the $\delta$ rest frame and so $\mt=m$. However, any boost with both transverse and longitudinal components increases $|\Delta\eta|$ and hence $\mtp<\mt$.

\item {\bf $\phi=\pi/2$ and $\theta\neq\pi/2$, any $\gamma$:} In this case $\Delta\eta\neq 0$ in the $\delta$ rest frame and any combined transverse and longitudinal boost decreases $|\Delta\eta|$ and increases $\mtp$. For this reason $\mt<\mtp\leq m$. Application of such a boost can therefore `regenerate' the Jacobian peak of events at the transverse mass end-point. We shall discuss this further below (Section~\ref{sec5}).

\item {\bf $\gamma> 1$, $\phi=0$ or $\phi=\pi$, any $\theta$:} In this case the decay axis lies in the plane of the beam direction and the boost. As is the case when the boost is purely transverse and $\phi=0$ or $\phi=\pi$, the direction of the momentum of $\chi$ for $\phi=0$ (or $v$ for $\phi=\pi$) in the transverse plane can change sign and become parallel to that of $v$ (or $\chi$). At this point the transverse mass becomes identically zero (remembering that $v$ and $\chi$ are massless). The criterion can be found by setting $\phi=0$ and solving the equation $p_{\chi, \mathrm{x}}'=0$, or equivalently $\mtp=0$, for $\gamma$ (and similarly for $p_{v, \mathrm{x}}'=0$ when $\phi=\pi$). The result is:
\begin{equation}
\gamma = \frac{-(\tan\xi\sin\theta+\cos\theta)\pm \sec^2\xi}{\tan\xi(\tan\xi\cos\theta-\sin\theta)}
\label{eqn:gamlim}
\end{equation}
When $\theta=\pi/2$ this reduces further to $\gamma=1\pm(\tan\xi+\cot\xi)$. This curve is visible in Figure~\ref{fig:gamxi}~(top) as the boundary of the dark blue region where $\mtp/m=0$. The `horns' in this figure lying at $\xi=0$, $\pm\pi/2$, and $\pm\pi$ correspond to cases where $\beta_{\mathrm{x}}=0$ or $\beta_{\mathrm{z}}=0$. In the former case the boost is purely longitudinal and hence $\mtp=\mt$ independently of the relativistic boost factor $\gamma$. In the latter case the boost is purely transverse and $\mtp=m$ independently of $\gamma$, as noted in the second bullet in Section~\ref{sec32}.

\item {\bf $\gamma\rightarrow\infty$:} In this limit $\mtp$ reduces to a simple form:
\begin{eqnarray}
\mtp &=& \frac{m p_{v, \mathrm{y}}}{\sqrt{E_v^2-(\bm\beta\cdot\bm p_v)^2}}\cr
&=& \frac{m\sin\theta\sin\phi}{\sqrt{1-\left(\sin\xi\sin\theta\cos\phi+\cos\xi\cos\theta\right)^2}}.
\label{eq:gaminfty}
\end{eqnarray}
When $\xi=\pi/2$, and hence $\beta_{\mathrm{x}}=1$, Eqn.~(\ref{eq:gaminfty}) reduces to Eqn.~(\ref{eqn:lim}), as expected. This function is shown in Figure~\ref{fig:gaminfty} (top) for the case $\theta=0.9$ and $\phi=0.2$. The peaks in $\mtp$ occur when:\\
\noindent\begin{minipage}{.5\linewidth}
\begin{eqnarray}
\beta_{\mathrm{x}} &=& \frac{\sin\theta\cos\phi}{\sqrt{1-\sin^2\theta\sin^2\phi}}\cr
&=& \frac{p_{v, \mathrm{x}}}{\sqrt{p_{v, \mathrm{x}}^2+p_{v, \mathrm{z}}^2}},\nonumber
\end{eqnarray}
\end{minipage}%
\begin{minipage}{.5\linewidth}
\begin{eqnarray}
\beta_{\mathrm{z}} &=& \frac{\cos\theta}{\sqrt{1-\sin^2\theta\sin^2\phi}}\cr
&=& \frac{p_{v, \mathrm{z}}}{\sqrt{p_{v, \mathrm{x}}^2+p_{v, \mathrm{z}}^2}},
\label{eq:mtmax}
\end{eqnarray}
\end{minipage}

\vspace{3mm}
and hence $\tan\xi = \tan\theta\cos\phi$. At these points 
\begin{eqnarray}
p_{v, \mathrm{x}}' &=& p_0\gamma\sin\theta\cos\phi\left(1+\frac{1}{\sqrt{1-\sin^2\theta\sin^2\phi}}\right) \gg p_{v, \mathrm{y}}',\cr
p_{v, \mathrm{z}}' &=& p_0\gamma\cos\theta\left(1+\frac{1}{\sqrt{1-\sin^2\theta\sin^2\phi}}\right),\cr
p_{\chi, \mathrm{x}}' &=& -p_0\gamma\sin\theta\cos\phi\left(1-\frac{1}{\sqrt{1-\sin^2\theta\sin^2\phi}}\right) \gg p_{\chi, \mathrm{y}}',\cr
p_{\chi, \mathrm{z}}' &=& -p_0\gamma\cos\theta\left(1-\frac{1}{\sqrt{1-\sin^2\theta\sin^2\phi}}\right).
\label{eqn:peakmom}
\end{eqnarray}
Consequently $\beta_{\mathrm{x}}/\beta_{\mathrm{z}}= p_{v, \mathrm{x}}'/p_{v, \mathrm{z}}'=p_{v, \mathrm{x}}/p_{v, \mathrm{z}}$, and the boost is hence parallel to the momentum of $v$ or $\chi$ in the $\hat{\mathrm{x}}-\hat{\mathrm{z}}$ plane. The transverse mass in the laboratory frame $\mtp$ equals $m$ and hence saturates the physical bound and can exceed the value of $\mt$ in the $\delta$ rest frame. We shall discuss the reason for this behaviour in Section~\ref{sec5}. The minimum value of $\mtp$ for $\gamma\rightarrow\infty$ occurs when:\\
\noindent\begin{minipage}{.5\linewidth}
\begin{eqnarray}
\beta_{\mathrm{x}} &=& \frac{\cos\theta}{\sqrt{1-\sin^2\theta\sin^2\phi}}\cr
&=& \frac{p_{v, \mathrm{z}}}{\sqrt{p_{v, \mathrm{x}}^2+p_{v, \mathrm{z}}^2}}, \nonumber
\end{eqnarray}
\end{minipage}%
\begin{minipage}{.5\linewidth}
\begin{eqnarray}
\beta_{\mathrm{z}} &=& \frac{\sin\theta\cos\phi}{\sqrt{1-\sin^2\theta\sin^2\phi}}\cr
&=& \frac{p_{v, \mathrm{x}}}{\sqrt{p_{v, \mathrm{x}}^2+p_{v, \mathrm{z}}^2}},
\label{eq:mtmin}
\end{eqnarray}
\end{minipage}
\vspace{3mm}

\noindent and hence $\tan\xi = \cot\theta/\cos\phi$. At these points $\mtp=\sin\theta\sin\phi$. 

\end{itemize}

\section{Discussion}
\label{sec5}

It is clear from the analysis of the dependence of $\mtp$ on $\gamma$ and $\xi$ presented in Section~\ref{sec4} that a combined transverse and longitudinal boost of $\delta$ can lead to transverse mass values which exceed those measured in the rest frame of $\delta$, and indeed can saturate the bound $\mtp\le m$. This observation is in stark contrast to the case of purely transverse boosts, where $\mtp\le\mt<m$ when $\theta\neq\pi/2$, as noted in Section~\ref{sec3}. In order to understand this we can consider the  special case where $\gamma\rightarrow\infty$ discussed in Section~\ref{sec42}. At the peaks in Figure~\ref{fig:gaminfty} (top), where $\mtp=m$, the ratios of $\hat{\mathrm{x}}$ and $\hat{\mathrm{z}}$ momentum components observed in the laboratory frame for both $v$ and $\chi$ (Eqn.~(\ref{eqn:peakmom})) are equal: $p_{v, \mathrm{x}}'/p_{v, \mathrm{z}}' = p_{\chi, \mathrm{x}}'/p_{\chi, \mathrm{z}}' = \tan\theta\cos\phi$. As $\gamma\rightarrow\infty$ we can neglect the $\hat{\mathrm{y}}$ components of the momenta of these particles and proceed to calculate $\theta'$, the value of $\theta$ observed in the laboratory frame: $\tan\theta' = p_{\mathrm{T}}'/p_{\mathrm{z}}' \simeq p_{\mathrm{x}}'/p_{\mathrm{z}}' = \tan\theta\cos\phi$ for both $v$ and $\chi$. The equivalence of $\theta'$ for both $v$ and $\chi$ causes their rapidities $\eta$ to be identical and hence $\Delta\eta=0$ and $\mtp=m$. In general this condition is satisfied along a curve in the $\gamma-\xi$ plane for specific values of $\theta$ and $\phi$, leading to a `ridge' with $\mtp=m$ in Figure~\ref{fig:gamxi} (bottom) and the peaks with $\mtp=m$ in Figure~\ref{fig:gaminfty} (top). 

When comparing Figure~\ref{fig:gamxi} (bottom) with Figure~\ref{fig:gaminfty} (top) two points may be noted:
\begin{itemize}
\item Two of the peaks in $\mtp$ expected at large values of $\gamma$ are not present when $\gamma\rightarrow\infty$. These are the peaks in Figure~\ref{fig:gamxi} (bottom) located near $\xi=0$ and $\xi=\pm\pi$, where $\beta_{\mathrm{x}}\sim0$, which are not visible in Figure~\ref{fig:gaminfty} (top).
\item When $\xi$ is exactly zero, $\beta_{\mathrm{x}}=0$ and hence the boost is purely longitudinal. We therefore expect $\mtp=\mt=m\sin\theta$. However, Eqn.~(\ref{eq:gaminfty}) indicates that when $\gamma\rightarrow\infty$ and $\beta_{\mathrm{x}}=0$, $\mtp=m\sin\phi$. Since we are, in general, free to pick events with arbitrary $\theta$ and $\phi$ these values are apparently in conflict.
\end{itemize}
The answer to these two conundrums is illustrated in Figure~\ref{fig:gaminfty} (bottom), which shows $\mtp/m$ over a narrow range of $\xi$ around $\xi=0$, both for $\gamma\rightarrow\infty$ and for several large, but finite, values of $\gamma$. At finite $\gamma$, the peak in $\mtp$ near $\xi=0$, obtained when $\Delta\eta=0$, narrows and moves towards $\xi=0$ as $\gamma$ becomes larger, with the function maintaining a value $\mtp=m\sin\theta$ when crossing the $\xi=0$ axis and $\mtp=m$ at the peak. When $\gamma\rightarrow\infty$ this peak becomes infinitely narrow and only in this unphysical case does the `underlying' value $\mtp=m\sin\phi$ then take over. 

One can consider whether this analysis can be used to reduce the dependence of the shape of the experimentally measured transverse mass distribution on the recoil of the parent particle under consideration. In particular, the selection of a population of events with transverse mass values invariant under the Lorentz boosts generated by this recoil would minimise the systematic uncertainties in the measurement of the mass of the parent particle arising from the modelling of the recoil. This would be very useful for measurements of, for instance, the mass of the $W$ boson. In the presence of purely transverse Lorentz boosts a requirement that the transverse momentum components of the decay products in the lab frame parallel to the boost direction be equal would ensure that $\phi=\pi/2$ and hence that events are indeed invariant under these boosts, as discussed in bullet 3 of Section~\ref{sec32}. Unfortunately however, this strategy fails in the presence of a combined transverse and longitudinal Lorentz boost, as discussed in Section~\ref{sec4}. In this case, in order to isolate events at the transverse mass end-point unaffected by the boost one would need to measure the rapidity of both the visible and invisible decay products of the parent. The latter is inaccessible experimentally and hence this strategy cannot be applied in practice. 

\section{Conclusions}
\label{sec6}

In this paper we have studied the transformation properties of the transverse mass under Lorentz boosts with a component transverse to the beam direction. When a purely transverse boost is applied, we observed that the transverse mass is invariant when (a) the momenta of the decay products of the parent particle are confined to the transverse plane and/or (b) the transverse momenta of these decay products are perpendicular to the boost direction. In cases where the boost is purely transverse, the transverse mass in the laboratory frame cannot exceed that in the rest frame of the parent particle. Turning to cases with a longitudinal component to the boost in addition to a transverse component, we observed that by contrast the transverse mass in the laboratory frame can exceed that in the parent's rest frame, and indeed can saturate the bound provided by the rest mass of the parent. In such cases the boost is such that the difference in rapidity between the momenta of the decay products in the laboratory frame vanishes and hence the transverse mass in this frame equals the invariant mass.

\section*{Acknowledgements}
This project has received funding from the European Research Council (ERC) under the European Union's Horizon 2020 research and innovation programme (grant agreement n$^{\mathrm{o}}$ 694202). DRT also acknowledges financial support from the UK Science and Technology Facilities Council (STFC).


%
%

\end{document}